\documentclass[symmetry,article,accept,oneauthor,pdftex]{Definitions/mdpi} 

\firstpage{1} 
\pubvolume{0}
\issuenum{1}
\articlenumber{5}
\pubyear{2020}
\copyrightyear{2020}
\history{Received: date; Accepted: date; Published: date}
\updates{yes} % If there is an update available, un-comment this line

\usepackage{bm}
\usepackage[eulergreek]{sansmath}
\newcommand{\mymathsf}[1]{\mbox{\sansmath$\mathsf{#1}$}}

\DeclareFontEncoding{LGR}{}{}
\DeclareSymbolFont{sfgreek}{LGR}{cmss}{m}{n}
\DeclareMathSymbol{\sfgamma}{\mathord}{sfgreek}{`g}
\DeclareMathSymbol{\sfpi}{\mathord}{sfgreek}{`p}

\usepackage[normalem]{ulem}

\Title{Combining 3-Momentum and Kinetic Energy on Galilei/Newton Spacetime%Attention AE/ME. The following layout issues have not been checked by the English Editing Department and must be carefully verified by the AE/Layout Department: All callout issues, bold usage of callouts, and references to callouts in the text. Correct callout usage in figures. Figure and Table layout issues. Footnote formatting and Glossaries have not been checked. En dash usage for negative values, en dash usage to indicate relationships, en dash usage to indicate bonds (especially in chemistry). The English Editing Department is not responsible for correct italic usage for genes, proteins and technical terminology. This responsibility belongs to the authors. The following are also not checked: spacing between numbers and units of measurement, ratios, en dashes for ranges, date and time formats, punctuation in equation lines, and less than/more than spacing (< >). Finally, capitalization and layout of titles/headings must be properly checked and ensuring 'Eq.' and 'Fig.' are properly spelled out, as these are layout issues.
}

\Author{{Christian Y. Cardall} $^{\dagger}$\orcidA{}}

\AuthorNames{Christian Y. Cardall}

\address{%
\quad Physics Division, Oak Ridge National Laboratory, Oak Ridge, TN 37831-6354, USA; cardallcy@ornl.gov}

\firstnote{\hangafter=1 \hangindent=1.05em \hspace{-0.82em} \textls[-15]{This manuscript has been authored by UT-Battelle, LLC, under contract DE-AC05-00OR22725 with the \mbox{US Department} of Energy (DOE). The US government retains and the publisher, by accepting the article \mbox{for publication}, acknowledges that the US government retains a nonexclusive, paid-up, irrevocable, \mbox{worldwide license} to publish or reproduce the published form of this manuscript, or allow others to do \mbox{so, for US government} purposes. DOE will provide public access to these results of federally sponsored research \mbox{in accordance} with the DOE {Public Access Plan} (\url{http://energy.gov/downloads/doe-public-access-plan})}.}%mdpi:website is not allowed, please delete it

\abstract{Without the mass-energy equivalence available on Minkowski spacetime $\mathbb{M}$, it is not possible on 4-dimensional non-relativistic Galilei/Newton spacetime $\mathbb{G}$ to combine 3-momentum and {total} mass-energy in a single tensor object.
However, given a fiducial frame, it {is} possible to combine 3-momentum and {kinetic} energy into a linear form (particle) or $(1,1)$ tensor (continuum) in a manner that exhibits increased unity of classical mechanics on flat relativistic and non-relativistic spacetimes $\mathbb{M}$ and $\mathbb{G}$. 
As on $\mathbb{M}$, for a material continuum on $\mathbb{G}$, the first law of thermodynamics can be considered a consequence of a unified dynamical law for energy-momentum rather than an independent postulate.
}


\begin{document}
%%%%%%%%%%%%%%%%%%%%%%%%%%%%%%%%%%%%%%%%%%

%%%%%%%%%%%
\section{Introduction}
\label{sec:Introduction}

Traditional points of departure for non-relativistic and relativistic classical mechanics (e.g., \cite{Landau1987Fluid-Mechanics,Ferrarese2008Introduction-to,Gourgoulhon2013Special-Relativ}) feature distinct pictures of space and time.
The traditional non-relativistic picture is that tensor fields on 3-dimensional Euclidean position space $\mathbb{E}^3$ evolve as functions of absolute time $t$. 
In contrast, underlying the relativistic picture is a unified 4-dimensional spacetime; for present purposes let this be flat Minkowski spacetime $\mathbb{M}$.
Tensor fields on $\mathbb{M}$ embody the history of the system---a kind of static, eternal reality governed by equations more in the character of constraints than evolution.
%Conjugate to unified spacetime is a single geometric object, 4-momentum, combining the energy and 3-momentum that are separate in the non-relativistic case. 
\mbox{On $\mathbb{M}$,} the 3-momentum and energy that are separate in the non-relativistic case are combined in a single geometric object, the 4-momentum.
 
 The attitude towards the traditional non-relativistic and relativistic formulations of classical mechanics is often limited to deriving the former as a limit of the latter.
As 4-dimensional equations are split into $3+1$ dimensions, unified balance of 4-momentum on $\mathbb{M}$ is decomposed into balance of 3-momentum and balance of energy.
Then the $c \rightarrow \infty$ (infinite speed of light) limit of the relativistic equations in $3+1$ dimensions coincides with the non-relativistic equations.

It is intriguing to consider the extent to which the conceptual relationship can be exploited in the reverse direction, by reformulating non-relativistic physics in light of the relativistic perspective: {Can the non-relativistic evolution equations on position space also be understood as constraint equations \mbox{on spacetime?}}

\textls[-25]{The answer has been yes, to a certain extent.
At least by the 1920s, Weyl \cite{Weyl1922Space---Time---} and Cartan \cite{Cartan1923Sur-les-variete,Cartan1924Sur-les-variete,Cartan1986On-manifolds-wi} considered the combination of Euclidean position space $\mathbb{E}^3$ and Euclidean absolute time $\mathbb{E}$ into a non-relativistic 4-dimensional spacetime.
Works by Toupin and Truesdell \cite{Toupin1957World-invariant,Truesdell1960The-Classical-F}, \mbox{Trautmann \cite{Trautman1965Foundations-and,Trautman1966Comparison-of-N}}, \mbox{and K\"untzle \cite{Kunzle1972Galilei-and-Lor}} bear mention as example entry points into what has been a rather mathematically-oriented literature across the intervening decades.
A recent work of \mbox{mine \cite{Cardall2019Minkowski-and-G}} compares---with additional discussion and references---Minkowski spacetime $\mathbb{M}$ and what I call Galilei/Newton spacetime $\mathbb{G}$, both of which are flat 4-dimensional manifolds, and indeed 4-dimensional affine spaces.
That work illustrates that kinetic theory on spacetime provides an intuitive understanding of fluid dynamics from a mostly 4-dimensional perspective on both relativistic and non-relativistic spacetimes, expressed by the same 4-dimensional equations for the fluxes of baryon number, 3-momentum, and kinetic+internal energy on both $\mathbb{M}$ and $\mathbb{G}$.}

However, a perceived inability to unite 3-momentum and energy in a single tensor in the non-relativistic case has remained notable.
In the 1970s Duval and K\"unzle \cite{Duval1978Dynamics-of-con} used a variational principle to derive for a non-relativistic material continuum a tensor unifying stresses and internal energy flux, but it excludes mass flow; its 4-divergence does not vanish in the absence of an external 4-force, but is equated to a bulk acceleration term.
More recently, de~Saxc\'e and Vall\'ee constructed a tensor of vanishing divergence that includes the non-relativistic kinetic energy of bulk motion, and exhibits the transformation properties of the latter under Galilei boosts, by assembling an ``energy-momentum-mass-tensor'' on a 5-dimensional extended non-relativistic spacetime \cite{de-Saxce2012Bargmann-group-,de-Saxce2016Galilean-Mechan,de-Saxce20175-Dimensional-T}. 
The extra dimension, and the associated transformation of this object in part under the Bargmann group (a central extension of the Galilei group), are necessary to capture the transformation properties of non-relativistic kinetic energy.

Here I present a tensor equation on 4-dimensional spacetime that encompasses both the 3-momentum and kinetic+internal energy of a material continuum; consists of a vanishing divergence in the absence of external 4-force per baryon; and does so in a conceptually unified way in both the relativistic and non-relativistic cases, that is, on both $\mathbb{M}$ and $\mathbb{G}$.
This is an equation for the divergence of what I call the {``relative energy-momentum flux tensor''} $\bm{S}$ (the adjective ``relative'' betrays the fact that it is defined in terms of a fixed family of fiducial frames, in such a way that the transformation properties of kinetic energy are not manifest). 
A key point is that $\bm{S}$ is a $(1,1)$ tensor, with components ${S^\mu}_\nu$, satisfying a linear form equation---in contrast to the $(2,0)$ {``total inertia-momentum flux tensor''} (a.k.a. energy-momentum or stress-energy tensor) $\bm{T}$, with components $T^{\mu\nu}$, satisfying a vector equation.
What happens is that Galilei invariance forbids energy contributions to 4-momentum and a time component of 4-force when these are considered as vectors.
However, this restriction does not apply to 4-momentum and 4-force regarded as linear forms.
%This is where internal energy density and flux have been hiding on $\mathbb{G}$ all along.
In this work I motivate $\bm{S}$ and its governing equation on both $\mathbb{M}$ and $\mathbb{G}$.

{The approach taken here is to use relativistic mechanics on $\mathbb{M}$ with its natural 4-dimensional perspective as a starting point, and discover what 4-dimensional perspective can make sense on non-relativistic spacetime $\mathbb{G}$ by considering what happens as $c \rightarrow \infty$.
As the speed of light $c$ ultimately plays no role in non-relativistic mechanics, there is irony in the fact that it plays a role in constructing a non-relativistic approach, in the negative sense that the tensors on $\mathbb{G}$ that emerge from the exercise can only be those from which $c$ disappears.
Such can be the nature of backporting insights from a superseding theory into the one it supersedes.
}

%%%%%%%%%%%%%%%%%%%%%%%%%%%%%%%%%
\section{Relativistic Classical Mechanics on Minkowski Spacetime}

Recall some aspects of the geometry of Minkowski spacetime $\mathbb{M}$ before outlining geometric treatments of classical particle and continuum mechanics thereon.

\subsection{Minkowski Spacetime $\mathbb{M}$}

On Minkowski spacetime $\mathbb{M}$ the history of a particle of mass $m$ is a worldline $\mathcal{X}(\tau)$ parametrized by the proper time $\tau$ measured by a clock riding along with the particle.
The tangent vector to the worldline, $\bm{U} = \mathrm{d}\mathcal{X}(\tau) / \mathrm{d}\tau$, is the 4-velocity of such a ``{comoving observer}.''
It satisfies the normalization $\bm{g}(\bm{U},\bm{U}) = -c^2$, where $\bm{g}$ is the spacetime metric.
The components of $\bm{g}$ and its inverse $\overleftrightarrow{\bm{g}}$ are gathered by the $3+1$ block matrices
\begin{equation}
\mathsf{g} = \begin{pmatrix} - c^2 & \mymathsf{0} \\  \mymathsf{0}  & \mymathsf{1} \end{pmatrix}, \ \ \ 
\overleftrightarrow{\mathsf{g}} = \mathsf{g}^{-1} = \begin{pmatrix} - 1 / c^2 & \mymathsf{0} \\  \mymathsf{0}  & \mymathsf{1} \end{pmatrix}
\label{eq:MetricRepresentation_M}
\end{equation} 
in any inertial frame.

Select a fiducial frame, a global inertial frame on $\mathbb{M}$.
Associated with this frame is a family of ``{fiducial observers}'' whose uniform 4-velocity field $\bm{w} = \partial  / \partial t$ (also normalized as $\bm{g}(\bm{w},\bm{w}) = -c^2$) is the tangent vector field to the coordinate lines of the global time coordinate $t$.
The level surfaces $\mathbb{S}_t$ of the global time coordinate $t$ foliate $\mathbb{M}$ into affine hyperplanes; these represent 3-dimensional position space with Euclid geometry embodied by a 3-metric $\bm{\gamma}$, and are associated with a uniform linear form $\bm{t} = \mathbf{d}t = \bm{\nabla}t$.
With the fiducial frame components of $\bm{w}$ and $\bm{t}$ gathered by 4-column and 4-row 
\begin{equation}
\mathsf{w} = \begin{pmatrix} 1 \\ \mymathsf{0} \end{pmatrix}, \ \ \ 
\mathsf{t} = \begin{pmatrix} 1 & \mymathsf{0} \end{pmatrix}
\label{eq:TimeFormRepresentation}
\end{equation}
respectively (in $3+1$ block form), it is clear from Equation~(\ref{eq:MetricRepresentation_M}) that $\bm{t}$ is related to $\bm{w}$ by
\begin{equation}
\bm{t} = -\frac{1}{c^2}\, \bm{g} \cdot \bm{w} = -\frac{1}{c^2}\, \underline{\bm{w}}.
\label{eq:TimeForm} 
\end{equation}
In this work the dot operator ($\cdot$) introduced in Equation~(\ref{eq:TimeForm})---which reads $t_\mu = - g_{\mu\alpha} \, w^\alpha / c^2 = - w_\mu / c^2$ in components---never denotes a scalar product of vectors, but only contraction with the first available index.
Instead, a scalar product of vectors will always be expressly given in terms of a metric tensor, for instance $\bm{g} (\bm{w},\bm{w})$ or $\bm{\gamma} (\bm{v},\bm{v})$.
The linear form $\underline{\bm{w}}$ is the index-lowered metric dual of $\bm{w}$, \mbox{with components} $w_\mu = g_{\mu\alpha} \, w^\alpha$.

The 3-metric $\bm{\gamma}$ and its index-raised siblings $\overleftarrow{\bm{\gamma}}$, $\overrightarrow{\bm{\gamma}}$and $\overleftrightarrow{\bm{\gamma}}$ on $\mathbb{S}_t$, with components $\gamma_{ij}$, ${\gamma^j}_i = {\gamma_i}^j = \delta_i^j$, and $\gamma^{ij}$ respectively, can also be regarded as tensors on $\mathbb{M}$ (with components $\gamma_{\mu\nu}$, ${\gamma^\nu}_\mu$, ${\gamma_\mu}^\nu$, and $\gamma^{\mu\nu}$) that behave as projection tensors to $\mathbb{S}_t$. 
(Warning: this convention on over-arrows for index-raising differs from that in \cite{Cardall2019Minkowski-and-G}). 
These can be expressed in terms of the metric tensor $\bm{g}$, \mbox{inverse $\overleftrightarrow{\bm{g}}$}, \mbox{and identity} tensor $\bm{\delta} = \overleftarrow{\bm{g}} = \overrightarrow{\bm{g}}$ on $\mathbb{M}$:
\begin{eqnarray}
\bm{\gamma} &=& \bm{g} + \frac{1}{c^2} \, \underline{\bm{w}} \otimes \underline{\bm{w}}, \ \ \ 
\overleftrightarrow{\bm{\gamma}} = \overleftrightarrow{\bm{g}} + \frac{1}{c^2} \, \bm{w} \otimes \bm{w}, 
\label{eq:Projection_W_1} \\
\overleftarrow{\bm{\gamma}} &=& \bm{\delta} + \frac{1}{c^2} \, \bm{w} \otimes \underline{\bm{w}}, \ \ \ 
\overrightarrow{\bm{\gamma}} = \bm{\delta} + \frac{1}{c^2} \, \underline{\bm{w}} \otimes \bm{w}.
\label{eq:Projection_W_2}
\end{eqnarray}
The components of these tensors all transform differently under Lorentz transformations.
However, in the fiducial inertial frame the single matrix
\begin{equation}
\sfgamma = \overleftarrow{\sfgamma} = \overrightarrow{\sfgamma} = \overleftrightarrow{\sfgamma}
= \begin{pmatrix} 0 & \mymathsf{0} \\ \mymathsf{0} & \mymathsf{1} \end{pmatrix}
\label{eq:ProjectionMatrix}
\end{equation}
gathers the components of all of them.

As the projection tensors to $\mathbb{S}_t$ have vanishing contractions with $\bm{w}$ and/or $\underline{\bm{w}}$, they can be used to decompose tensors on $\mathbb{M}$ into spacelike and timelike pieces. 
For instance, the 4-velocity $\bm{U}$ decomposes as 
\begin{equation}
\bm{U} = \Lambda_{\bm{v}} \left( \bm{w} + \bm{v} \right), 
\label{eq:FourVelocity_M}
\end{equation}
in which $\Lambda_{\bm{v}} = \left(1 - \bm{\gamma}(\bm{v},\bm{v}) / c^2 \right)^{-1/2}$ is the Lorentz factor following from the normalization, and the 3-velocity $\bm{v}$ defined from the projection $\Lambda_{\bm{v}}  \bm{v} = \overleftarrow{\bm{\gamma}} \cdot \bm{U} = \bm{U} \cdot \overrightarrow{\bm{\gamma}}$ is a vector on $\mathbb{M}$ that happens to be tangent to $\mathbb{S}_t$ (and therefore as needed could also be regarded as simply a vector on $\mathbb{S}_t$).

We will also need decompositions relative to $\bm{U}$. 
Such decompositions allow specification of material properties by defining quantities measured by a comoving observer.
They are accomplished at any point $\mathcal{X}$ of $\mathbb{M}$ with the projection tensors 
\begin{eqnarray}
\bm{h} &=& \bm{g} + \frac{1}{c^2} \, \underline{\bm{U}} \otimes \underline{\bm{U}}, \ \ \ 
\overleftrightarrow{\bm{h}} = \overleftrightarrow{\bm{g}} + \frac{1}{c^2} \, \bm{U} \otimes \bm{U}, 
\label{eq:Projection_U_1} \\
\overleftarrow{\bm{h}} &=& \bm{\delta} + \frac{1}{c^2} \, \bm{U} \otimes \underline{\bm{U}}, \ \ \ 
\overrightarrow{\bm{h}} = \bm{\delta} + \frac{1}{c^2} \, \underline{\bm{U}} \otimes \bm{U}
\label{eq:Projection_U_2}
\end{eqnarray}
to a local hyperplane $\mathbb{S}_{\bm{U}(\mathcal{X})}$ that is the orthogonal complement of $\bm{U}(\mathcal{X})$.

{
Note also projections of the spacetime covariant derivative operator $\bm{\nabla}$. 
We have
\begin{eqnarray}
\frac{\partial(\ )}{\partial t} &=& \bm{w} \cdot \bm{\nabla}(\ ), \\
 \bm{D}(\ )& =& \overrightarrow{\bm{\gamma}} \cdot \bm{\nabla} (\ )
\end{eqnarray}
for the projections relative to $\bm{w}$ of the spacetime covariant derivative on $\mathbb{M}$, where $\bm{D}$ is the covariant derivative on $\mathbb{S}_t$ associated with its Euclidean 3-metric $\bm{\gamma}$.
Alternatively
\begin{eqnarray}
\frac{\mathrm{d}(\ )}{\mathrm{d}\tau} &=& \bm{U} \cdot \bm{\nabla} (\ ), 
\label{eq:DerivativeProperTime} \\
\bm{\mathcal{D}}(\ ) &=& \overrightarrow{\bm{h}} \cdot \bm{\nabla} (\ )
\label{eq:ComovingDerivative}
\end{eqnarray}
are the projections relative to $\bm{U}$.
}

%%%%%%%%%%%%
\subsection{Particle Mechanics on $\mathbb{M}$}
\label{sec:Particles_M}

Consider now the dynamics of a relativistic material particle.
%\DELETE{For purposes of relating to the non-relativistic case, take care to distinguish between the {inertia-momentum vector} $\bm{I} = m \, \bm{U}$ of a particle and its {total energy-momentum form}, the metric dual $\underline{\bm{I}} = m \, \underline{\bm{U}}$, where $\underline{\bm{U}} = \bm{g} \cdot \bm{U}$. 
%The timelike components of the 4-column and 4-row} 
%\begin{equation}
%\DELETE{ \mathsf{I} = \begin{pmatrix} m \, \Lambda_\mathsf{v} \\ m \, \Lambda_\mathsf{v} \, \mathsf{v} \end{pmatrix}, \ \ \ 
%\underline{\,\mathsf{I}\,} = \begin{pmatrix} - m c^2 \, \Lambda_\mathsf{v} & m \, \Lambda_\mathsf{v} \, \mathsf{v^T} \end{pmatrix} }
%\label{eq:InertiaEnergyMomentumRepresentation}
%\end{equation}
%\DELETE{gathering their components in the fiducial frame confirm the appropriateness of this nomenclature (see Equations~(\ref{eq:FourVelocity_M}), (\ref{eq:TimeFormRepresentation}), and (\ref{eq:MetricRepresentation_M}); $\mathsf{v}$ is the 3-column gathering the components of $\bm{v}$ on $\mathbb{S}_t$).}
{In relativistic mechanics the vector 
\begin{equation}
\bm{I} = m \, \bm{U}
\label{eq:InertiaMomentumVector}
\end{equation}
(or $I^\mu = m \, U^\mu$) and its metric dual, the linear form $\underline{\bm{I}} = m \, \underline{\bm{U}}$, where $\underline{\bm{U}} = \bm{g} \cdot \bm{U}$ (or $I_\mu = m \, U_\mu$, where $U_\mu = g_{\mu\alpha}\, U^\alpha$), are both known as the energy-momentum, or simply 4-momentum.
They are represented by a 4-column and 4-row respectively:}
\begin{equation}
{ \mathsf{I} = \begin{pmatrix} m \, \Lambda_\mathsf{v} \\ m \, \Lambda_\mathsf{v} \, \mathsf{v} \end{pmatrix}, \ \ \ 
\underline{\,\mathsf{I}\,} = \begin{pmatrix} - m c^2 \, \Lambda_\mathsf{v} & m \, \Lambda_\mathsf{v} \, \mathsf{v^T} \end{pmatrix} }
\label{eq:InertiaEnergyMomentumRepresentation}
\end{equation}%mdpi: we changed Equations (7), (2) and (1) to  (1), (2) and 7
{(see {Equations}~(\ref{eq:MetricRepresentation_M}), (\ref{eq:TimeFormRepresentation}), and (\ref{eq:FourVelocity_M}); $\mathsf{v}$ is the 3-column gathering the components of $\bm{v}$ on $\mathbb{S}_t$). 
The spatial parts of both of these consist of the 3-momentum, but their time components differ: the contravariant component $I^0 = m \, \Lambda_\mathsf{v}$ is the relativistic inertia, while the covariant component $I_0 =  - m c^2 \, \Lambda_\mathsf{v}$ is the (negative of) relativistic total energy, including rest mass.
This distinction between inertia and energy is a technicality in relativistic mechanics, in which metric duality unifies the two concepts.
\mbox{However, the} distinction remains important in non-relativistic mechanics, which certainly has inertia, \mbox{but in} which the concept of rest mass does not exist ($m c^2$ is nonsense as $c \rightarrow \infty$).
Therefore, in order to treat the relativistic and non-relativistic cases in parallel in Section~\ref{sec:Unified}, in this paper I use the non-standard nomenclature ``{inertia-momentum vector}'' for $\bm{I}$, which exists on both $\mathbb{M}$ and $\mathbb{G}$; and ``{total energy-momentum form}'' for $\underline{\bm{I}}$, which exists only on $\mathbb{M}$.
It will turn out in Section~\ref{sec:Unified} that what I call a ``{relative} energy-momentum form'' $\bm{P}$, which excludes rest mass, does not depend on $c$ and therefore {can} be defined on both $\mathbb{M}$ and $\mathbb{G}$.
}

%\DELETE{Linear}
{Returning to the relativistic case, linear} form and vector versions of Newton's second law for particles on $\mathbb{M}$ read
\begin{equation}
\frac{\mathrm{d}}{\mathrm{d}\tau} \, \underline{\bm{I}} = \bm{\Upsilon}, \ \ \ 
\frac{\mathrm{d}}{\mathrm{d}\tau} \, \bm{I} = \overrightarrow{\bm{\Upsilon}}
\label{eq:ParticleNewtonSecondLaw_M}
\end{equation}
in terms of the 4-force linear form $\bm{\Upsilon}$ and vector $\overrightarrow{\bm{\Upsilon}} = \bm{\Upsilon} \cdot \overleftrightarrow{\bm{g}}$ (that is, $\Upsilon^{\mu} = \Upsilon_\alpha \, g^{\alpha\mu}$).
The 4-force vector can be decomposed relative to either $\bm{U}$ or $\bm{w}$,
\begin{equation}
\overrightarrow{\bm{\Upsilon}} = \frac{\theta}{c^2} \, \bm{U} + \overrightarrow{\bm{f}}
= \frac{\Theta}{c^2} \, \bm{w} + \overrightarrow{\bm{F}},
\label{eq:FourForce_U_M}
\end{equation}
with heating rates per baryon and 3-force vectors $\theta, \overrightarrow{\bm{f}}$ or $\Theta, \overrightarrow{\bm{F}}$ measured by comoving or fiducial observers respectively, projected out by contraction with $\underline{\bm{U}}, \overrightarrow{\bm{h}}$ or $\underline{\bm{w}}, \overrightarrow{\bm{\gamma}}$.
Note that $\theta = - \underline{\bm{U}} \cdot \overrightarrow{\bm{\Upsilon}} = 0$ for an ``elementary'' particle of constant rest mass $m$.
Note that the 4-force $\bm{\Upsilon}$---and 3-forces $\bm{f}$ (according to a comoving observer) and $\bm{F}$ (according to a fiducial observer)---are here taken to be most naturally regarded as linear forms.
{Regarding force primarily as a linear form is more natural because it allows it to be contracted with displacement or velocity (vectors both) to yield work or power without the need for a metric.}

%%%%%%%%%

\subsection{Continuum Mechanics on $\mathbb{M}$}

{
\textls[-25]{Turning to a material continuum, its classical mechanics on $\mathbb{M}$ are governed by the spacetime constraints}
\begin{eqnarray}
\bm{\nabla} \cdot \bm{N} &=& 0,
\label{eq:BaryonConservation} \\
\bm{\nabla} \cdot \bm{T} &=& n \overrightarrow{\bm{\Upsilon}}
\label{eq:InertiaMomentumBalance}
\end{eqnarray}
on the baryon number flux vector $\bm{N}$ and total inertia-momentum flux tensor $\bm{T}$, a $(2,0)$ tensor, \mbox{where $\bm{\nabla}$} is the spacetime covariant derivative operator.
There is of course more than one way to arrive at these relativistic conservation laws.
One approach is to derive the mechanics of one kind of material continuum---a gas of classical particles---from relativistic kinetic theory (e.g., \cite{Cardall2019Minkowski-and-G}). 
This is a very direct and ``hands-on'' way of developing intuition for the physical meaning of the several scalars and tensors into which $\bm{N}$ and $\bm{T}$ can be decomposed. 
However, not all material continua are gases, and the kinetic theory of classical particles is not a fundamental physical theory.
In fact Equations~(\ref{eq:BaryonConservation}) and (\ref{eq:InertiaMomentumBalance}) do not depend on any particular microphysical model, and can be motivated on more general grounds (\mbox{e.g., \cite{Ferrarese2008Introduction-to}}).
Here I present a modified version of this latter approach, streamlined with \mbox{physical reasoning.}

The idea of a material continuum is that there is some connected region of measurable ``stuff,'' a quantity of matter, which can be neither created nor destroyed.
Understanding that mass is not conserved in a relativistic setting, take this quantity of matter to be baryon number instead.
\mbox{In mathematizing} the conservation of baryon number, Equation~(\ref{eq:BaryonConservation}) amounts to a description of the {kinematics} of the material continuum. 
As a balance equation for inertia-momentum (equivalent to energy-momentum on $\mathbb{M}$ if not on $\mathbb{G}$), Equation~(\ref{eq:InertiaMomentumBalance}) characterizes the {dynamics} of the material continuum: it is, for infinitesimal elements of the continuum, what the relativistic version of Newton's second law in Equation~(\ref{eq:ParticleNewtonSecondLaw_M}) is for point particles.

Consider what Equation~(\ref{eq:BaryonConservation}) signifies for infinitesimal elements of the material continuum, thereby relating the baryon number flux $\bm{N}$ to a 4-velocity field $\bm{U}$ of comoving observers and the baryon number density $n$ measured by those observers.
Let each element consist of a particular piece of material with fixed baryon number $\mathcal{N}$, confined to an infinitesimal region of 3-space whose spacetime position and 3-volume $\mathcal{V}$ vary with the flow of the continuum on spacetime, defined by $\bm{N}$.
As with the particles discussed in Section~\ref{sec:Particles_M}, each material element has a worldline $\mathcal{X}(\tau)$ with tangent vector $\bm{U}$ at each point.
Taken together, the wordlines of the continuum elements compose a congruence of curves filling $\mathbb{M}$, giving rise to a 4-velocity field $\bm{U}$ thereon.
This field is defined by its alignment with $\bm{N}$, that is,  
\begin{equation}
\bm{N} = n\, \bm{U}, 
\label{eq:ContinuumFourVelocity}
\end{equation}
where the scalar field $n = \mathcal{N} / \mathcal{V}$ is the baryon number density measured by an observer riding along with a fluid element (``comoving observer'').
Using Equations~(\ref{eq:DerivativeProperTime}) and (\ref{eq:ContinuumFourVelocity}), baryon number conservation as expressed by the vanishing divergence in Equation~(\ref{eq:BaryonConservation}) is equivalent to
\begin{equation}
\frac{1}{\mathcal{V}} \frac{\mathrm{d}\mathcal{V}}{\mathrm{d}\tau} = \bm{\nabla} \cdot \bm{U},
\end{equation}
relating the rate of change of a continuum element's 3-volume $\mathcal{V}$ along its worldline to the local spread of neighboring elements' worldlines, as represented by the divergence of $\bm{U}$.

Having defined a comoving velocity field $\bm{U}$ by the flow of conserved material (here taken to be baryon number), the flux of inertia-momentum $\bm{T}$ can be decomposed into quantities measured by the associated comoving observers---that is, with respect to the infinitesimal material elements, characterizing intrinsic material properties apart from their motion.

Begin consideration of the flux of inertia-momentum $\bm{T}$ by separating that portion of inertia-momentum which corresponds to the very existence of the matter (baryon number) constituting the material continuum:
\begin{equation}
\bm{T} = n \, \bm{U} \otimes \bm{I} - \overrightarrow{\bm{\Sigma}}.
\label{eq:InertiaMomentumTensor}
\end{equation}
Here $\bm{I} = m \, \bm{U}$ is the particle inertia-momentum vector introduced in Equation~(\ref{eq:InertiaMomentumVector}), with $m$ now interpreted as the mass per baryon.
(Separation of baryon rest mass is somewhat artificial in a relativistic context, but is physically insightful and will prove important in back-porting relativistic insights to the non-relativistic case. In practice the baryon mass $m$ may reflect a variable average in allowance of multiple nuclear species, or may be set to a reference constant, with energy associated with nuclear composition changes included in the internal energy density $\epsilon$ to be introduced shortly \cite{Cardall2017Relativistic-an}.)
In the first term of Equation~(\ref{eq:InertiaMomentumTensor}), the vector $\bm{I}$---the second factor of the tensor product---represents the intrinsically baryonic portion of inertia-momentum per baryon; meanwhile, the first factor of the tensor product represents the fact that this inertia-momentum per baryon moves in the direction $n \, \bm{U}$, being carried along as an intrinsic property of the material element itself.
Thanks to Equations~(\ref{eq:DerivativeProperTime}), (\ref{eq:InertiaMomentumVector}), (\ref{eq:BaryonConservation}), (\ref{eq:ContinuumFourVelocity}), and (\ref{eq:InertiaMomentumTensor}), the inertia-momentum balance expressed in Equation~(\ref{eq:InertiaMomentumBalance}) is equivalent to
\begin{equation}
n \, \frac{\mathrm{d}\bm{I}}{\mathrm{d}\tau}   = n \, \frac{\mathrm{d}}{\mathrm{d}\tau} \left( m \, \bm{U} \right) = n \overrightarrow{\bm{\Upsilon}} + \bm{\nabla} \cdot \overrightarrow{\bm{\Sigma}}.
\label{eq:ContinuumNewtonSecondLaw_M}
\end{equation}
This is Newton's second law for an infinitesimal element of the material continuum, \mbox{governing the} curvature of its 4-velocity $\bm{U}$.
In addition to the body 4-force $n \overrightarrow{\bm{\Upsilon}}$, the divergence $\bm{\nabla} \cdot \overrightarrow{\bm{\Sigma}}$ describes the net surface 4-force from neighboring continuum elements. 
If the 4-stress $\overrightarrow{\bm{\Sigma}}$ vanishes, Equation~(\ref{eq:ContinuumNewtonSecondLaw_M}) manifestly reduces to Equation~(\ref{eq:FourForce_U_M}), Newton's second law for a \mbox{relativistic particle.}

Note that the 4-stress $\overrightarrow{\bm{\Sigma}}$ appearing in Equation~(\ref{eq:InertiaMomentumTensor})---which distinguishes an infinitesimal element of a material continuum from a mere particle---has an index raised in order to include it in a $(2,0)$ tensor equation: an important point of this paper is that the 4-stress is to be regarded most naturally as a $(1,1)$ tensor, so that in Equation~(\ref{eq:InertiaMomentumTensor}) it appears as $\overrightarrow{\bm{\Sigma}} = \bm{\Sigma} \cdot \overleftrightarrow{\bm{g}}$ (that is, $\Sigma^{\mu\nu} = {\Sigma^\mu}_\alpha \, g^{\alpha\nu}$).
A {flux} is a ``vector-type'' (upper index) entity, while {inertia-momentum} and {energy-momentum} are respectively of ``vector-type'' (upper index) and ``linear-form-type'' (lower index), as discussed in Section~\ref{sec:Particles_M}. 
\mbox{Thus the} {flux of inertia-momentum} $\bm{T}$ is a $(2,0)$ tensor, while the {flux of energy-momentum} is a $(1,1)$ tensor, \mbox{as mentioned} in Section~\ref{sec:Introduction} and to be elaborated in Sections~\ref{sec:GalileiNewton} and \ref{sec:Unified}.

While Equations~(\ref{eq:InertiaMomentumBalance}), (\ref{eq:InertiaMomentumTensor}), and (\ref{eq:ContinuumNewtonSecondLaw_M}) provide an initial understanding of the relationship of inertia-momentum balance to the motion of an individual continuum element, further understanding of the 4-stress $\overrightarrow{\bm{\Sigma}}$ is needed.
This is obtained from the observation that an element of a material continuum is distinguished from a particle by the fact that, as a piece of matter with non-vanishing (if infinitesimal) extent, it is to be regarded as a tiny thermodynamic system in and of itself. 
Recognizing that all forms of energy contribute to inertia in relativity, and that the thermodynamic internal energy is (like baryon mass) an internal property of a continuum element carried along with its motion in the direction $n \, \bm{U}$,
continue the dissection of $\bm{T}$ by rewriting Equation~(\ref{eq:InertiaMomentumTensor}) as
\begin{eqnarray}
\bm{T} &=& n \, \bm{U} \otimes \bm{I} + n \, \bm{U} \otimes \frac{\epsilon}{c^2 n} \, \bm{U} - \overrightarrow{\bm{\Sigma}}'  
\label{eq:InertiaMomentumTensor_2} \\
 &=&  n \, \bm{U} \otimes \left( m + \frac{\epsilon}{c^2 n} \right) \bm{U} - \overrightarrow{\bm{\Sigma}}'. 
\label{eq:InertiaMomentumTensor_3}
\end{eqnarray}
Here $\epsilon$ is the internal energy density, so that the middle term of Equation~(\ref{eq:InertiaMomentumTensor_2}) separates from the 4-stress $\overrightarrow{\bm{\Sigma}}$ the contribution of a continuum element's internal energy to the flux of inertia-momentum inherent to its motion.
This completes isolation of the ``time-time'' portion of $\bm{T}$.

The remaining part of the 4-stress tensor describes another consequence of a continuum element's (infinitesimally) extended nature: the inertia-momentum that flows through the boundaries of the infinitesimal 4-volume swept out by the element's 3-volume $\mathcal{V}$ in time $\mathrm{d}\tau$.
The purely spatial part of $\overrightarrow{\bm{\Sigma}}$ (and $\overrightarrow{\bm{\Sigma}}$') is well-known as the 3-stress $\overrightarrow{\bm{\varsigma}}$, the 3-force exerted on the walls of a continuum element by neighboring elements, leaving
\begin{equation}
\overrightarrow{\bm{\Sigma}}' = n \, \bm{U} \otimes \frac{\epsilon}{c^2 n} \, \bm{U} - \overrightarrow{\bm{\Sigma}}''  - \overrightarrow{\bm{\varsigma}}.
\label{eq:InertiaMomentumTensor_4}
\end{equation}
With the ``time-time'' and ``space-space'' pieces accounted for, what remains in $\overrightarrow{\bm{\Sigma}}''$ are the ``time-space'' and ``space-time'' pieces.
Recall from the properties of relativistic spacetime (e.g., \mbox{in general} relativity) that $\bm{T}$ is symmetric; this is consistent with the manifest symmetry of the $\bm{U} \otimes \bm{U}$ term, and with the familiar fact that the 3-stress $\overrightarrow{\bm{\varsigma}}$ is symmetric (as a consequence of angular momentum conservation).
Recognizing that $\overrightarrow{\bm{\Sigma}}''$ must also be symmetric, and noting that we have not yet accounted for heat exchange between neighboring material elements, take
\begin{equation}
\overrightarrow{\bm{\Sigma}}'' = - \bm{U} \otimes \frac{\bm{q}}{c^2} - \frac{\bm{q}}{c^2} \otimes \bm{U},
\end{equation}
interpreting $\bm{q}$ as the 3-flux of internal energy out of a continuum element, orthogonal to $\bm{U}$.
\mbox{(Just as} angular momentum conservation and the resulting symmetry of $\overrightarrow{\bm{\varsigma}}$ follow from the rotational invariance of 3-space and the material continuum, so also invariance of $\mathbb{M}$ under ``pseudo-rotations'' or Lorentz boosts underlies the symmetry of $\overrightarrow{\bm{\Sigma}}''$.)
For the 4-stress as a whole, write
\begin{equation}
\overrightarrow{\bm{\Sigma}} = - n \, \bm{U} \otimes \frac{\epsilon}{c^2 n} \, \bm{U} - n \, \bm{U} \otimes \frac{\bm{q}}{c^2 n} - \bm{q} \otimes \frac{\bm{U}}{c^2}  + \overrightarrow{\bm{\varsigma}}.
\label{eq:FourStress_UU_U_M}
\end{equation}
The presence of two terms involving $\bm{q}$, forced upon us by the symmetry of the inertia-momentum tensor $\bm{T}$ and therefore the 4-stress tensor $\overrightarrow{\bm{\Sigma}}$, represents the dual role it plays in the relativistic case.
In one term (the third in Equation~(\ref{eq:FourStress_UU_U_M})), it represents internal energy flux.
In the other (appearing with factors of $n$ in the second term of Equation~(\ref{eq:FourStress_UU_U_M})), it constitutes an additional thermodynamic contribution to the 3-momentum carried by a continuum element in the direction $n\,\bm{U}$.
In connection with this last insight, it makes sense to define an ``internal energy flux vector'' $\bm{\Xi}$ by
\begin{equation}
\bm{\Xi} = \epsilon \, \bm{U} + \bm{q},
\end{equation}
in terms of which the 4-stress reads
\begin{equation}
\overrightarrow{\bm{\Sigma}} = - n \, \bm{U} \otimes \frac{\bm{\Xi}}{c^2 n}  - \bm{q} \otimes \frac{\bm{U}}{c^2}  + \overrightarrow{\bm{\varsigma}},
\end{equation}
and Equation~(\ref{eq:InertiaMomentumTensor}) for the inertia-momentum tensor becomes
\begin{equation}
\bm{T} = n \, \bm{U} \otimes \left( \bm{I}  + \frac{\bm{\Xi}}{c^2 n} \right) + \bm{q} \otimes \frac{\bm{U}}{c^2} - \overrightarrow{\bm{\varsigma}}.
\label{eq:InertiaMomentumTensor_5}
\end{equation}
In physical terms, this separates the inertia-momentum per baryon $\bm{I}  + \bm{\Xi} / c^2 n$ carried by the infinitesimal elements of the material continuum themselves from the heat and momentum fluxes passing through their surfaces, as represented by the last two terms in Equation~(\ref{eq:InertiaMomentumTensor_5}). 

One orthogonal decomposition of the 4-stress---that relative to comoving observers with 4-velocity $\bm{U}$---is given in Equation~(\ref{eq:FourStress_UU_U_M}), but it is not the only decomposition of potential interest.
The internal energy density $\epsilon$ and 3-flux $\bm{q}$, and the 3-stress $\bm{\varsigma}$, measured by a comoving observer can be projected out from $\overrightarrow{\bm{\Sigma}}$ by the appropriate contractions with $\underline{\bm{U}}$ and $\overrightarrow{\bm{h}}$.
Alternatively, $\overrightarrow{\bm{\Sigma}}$ can be decomposed as
\begin{equation}
\overrightarrow{\bm{\Sigma}} = - \bm{w} \otimes \frac{E}{c^2} \, \bm{w} - \bm{w} \otimes \frac{\bm{Q}}{c^2} - \frac{\bm{Q}}{c^2} \otimes \bm{w} + \overrightarrow{\bm{\sigma}} 
\label{eq:FourStress_UU_W_M}
\end{equation}
in terms of internal energy density $E$ and 3-flux $\bm{Q}$, and 3-stress $\bm{\sigma}$, measured by a fiducial observer and projected out by contractions with $\underline{\bm{w}}, \overrightarrow{\bm{\gamma}}$; these will include contributions from the bulk motion of the continuum relative to the fiducial observers.

Decomposed to $3+1$ dimensions and in the limit $c \rightarrow \infty$, Equations~(\ref{eq:BaryonConservation}) and (\ref{eq:InertiaMomentumBalance}) reduce to various non-relativistic formulations.
These are obtained with the decompositions in Equations~(\ref{eq:FourVelocity_M}), (\ref{eq:InertiaMomentumVector}), (\ref{eq:FourForce_U_M}), (\ref{eq:ContinuumFourVelocity}), (\ref{eq:InertiaMomentumTensor}), (\ref{eq:FourStress_UU_U_M}), (\ref{eq:FourStress_UU_W_M}), and projections along $\bm{w}$ or $\bm{U}$ and perpendicular to them using Equations~(\ref{eq:Projection_W_1}) and (\ref{eq:Projection_W_2}) or (\ref{eq:Projection_U_1}) and (\ref{eq:Projection_U_2}) respectively.
In the non-relativistic limit as $c \rightarrow \infty$,
\begin{eqnarray}
\frac{\mathrm{d}(\ )}{\mathrm{d}\tau}  &\rightarrow& \frac{\mathrm{d}(\ )}{\mathrm{d}t}, \\
\bm{\mathcal{D}}(\ )  &\rightarrow& \bm{D}(\ ) - \bm{t} \, (\bm{v} \cdot \bm{D})(\ ),
\end{eqnarray}
where
\begin{equation}
\frac{\mathrm{d}(\ )}{\mathrm{d}t} = \frac{\partial (\ )}{\partial t} + \bm{v} \cdot \bm{D} (\ )
\label{eq:MaterialDerivative}
\end{equation}
is the non-relativistic material derivative.
}

%%%%%%%%%%%%%%%%%%%%%%%%%%%%%%%%%%%%%%%
\section{Galilei/Newton Spacetime, Baryon Conservation, and Mass Conservation}
\label{sec:GalileiNewton}

While index raising and lowering via the spacetime metric preserves information on $\mathbb{M}$, \mbox{the geometry} of Galilei/Newton spacetime $\mathbb{G}$ provides only information-destroying projection operators to go from $(1,1)$ tensors to $(2,0)$ or $(0,2)$ tensors.
This has no major consequence for baryon conservation, which is governed by the spacetime divergence of the vector field $\bm{N}$.
However, on $\mathbb{G}$ the $(2,0)$ stress-inertia tensor $\bm{T}$ loses information about internal energy and stresses, so that a timelike projection of its divergence reduces to mass conservation rather than a full expression of energy conservation as on $\mathbb{M}$.
 
\subsection{Galilei/Newton Spacetime $\mathbb{G}$ and Its Geometric Consequences}

While Galilei/Newton spacetime $\mathbb{G}$---the non-relativistic analogue of Minkowski spacetime $\mathbb{M}$---has a qualitatively distinct geometric character, in many ways it can be understood as the $c \rightarrow \infty$ limit of the latter \cite{Cardall2019Minkowski-and-G}.
The absolute object on $\mathbb{M}$ governing causality is the metric $\bm{g}$ (with inverse $\overleftrightarrow{\bm{g}}$), which embodies the lightcones. 
As $c \rightarrow \infty$ these lightcones are ``pressed down'' into fixed spatial hyperplanes $\mathbb{S}_t$ with a unique linear form field $\bm{t}$ embodying absolute time.
A spacetime metric no longer makes sense (see Equation~(\ref{eq:MetricRepresentation_M}))---$\mathbb{G}$ is not a pseudo-Riemann manifold---but the inverse metric $\overleftrightarrow{\bm{g}}$ limits sensibly to the degenerate inverse ``metric'' $\overleftrightarrow{\bm{\gamma}}$ (compare {Equations}~(\ref{eq:MetricRepresentation_M}), (\ref{eq:Projection_W_1}) and (\ref{eq:ProjectionMatrix})), whose Galilei invariance is the remnant of Lorentz invariance that survives the limit.
The projection tensors $\bm{\gamma}$, $\overleftarrow{\bm{\gamma}}$, and $\overrightarrow{\bm{\gamma}}$ also exist (now regarded as separate tensors on $\mathbb{G}$ unrelated by spacetime metric duality), and the 4-vector field $\bm{w}$ has the same role associated with a fiducial inertial frame.
\mbox{While the} contraction $\bm{t} \cdot \bm{w} = 1$ still holds, the metric relationship $\overleftrightarrow{\bm{g}} \cdot \bm{t} = -c^{-2} \bm{w}$ on $\mathbb{M}$ degenerates to $ \overleftrightarrow{\bm{\gamma}} \cdot \bm{t} = \bm{0}$ on $\mathbb{G}$.

The congruence of worldlines of continuum elements exists on $\mathbb{G}$, with tangent vector field $\bm{U}$.
This 4-velocity is still related to baryon number flux by Equation~(\ref{eq:ContinuumFourVelocity}), but is now related to the 3-velocity by
\begin{equation}
\bm{U} = \bm{w} + \bm{v}.
\label{eq:FourVelocity_G}
\end{equation}
Baryon number conservation is still expressed by Equation~(\ref{eq:BaryonConservation}).

Even without the notions of a (pseudo-)norm or orthogonality afforded by a spacetime metric, spacelike projections and in particular decompositions with respect to timelike vector fields are still available on $\mathbb{G}$, but care must be taken to understand their geometric implications.

On $\mathbb{M}$ the tensors $\overleftarrow{\bm{\gamma}}$ and $\overleftarrow{\bm{h}}$ and their siblings project to spacelike hyperplanes $\mathbb{S}_t$ and $\mathbb{S}_{\bm{U}(\mathcal{X})}$ that are orthogonal to $\bm{w}$ and $\bm{U}$ respectively.
The corresponding availability of {orthogonal} decompositions allows some flexibility in how these are expressed---various combinations of up and down indices, without information loss.

On $\mathbb{G}$ the situation is different.
Absolute time means that $\mathbb{S}_t$ are the only spacelike hypersurfaces, and the degeneracy of  
$\overleftrightarrow{\bm{\gamma}}$ means that projections to $\mathbb{S}_t$ are not unique; $\overleftrightarrow{\bm{h}}$ and its siblings also project to $\mathbb{S}_t$.
While on $\mathbb{G}$ the tensors $\overleftarrow{\bm{\gamma}}$, $\overrightarrow{\bm{\gamma}}$ and $\overleftarrow{\bm{h}}$, $\overrightarrow{\bm{h}}$ with vanishing contractions with $\bm{w}$ and $\bm{U}$ respectively can be expressed 
\begin{eqnarray}
\overleftarrow{\bm{\gamma}} &=& \bm{\delta} - \bm{w} \otimes \bm{t}, \ \ \ 
\overrightarrow{\bm{\gamma}} = \bm{\delta} - \bm{t} \otimes \bm{w}, 
\label{eq:Projection_W_G} \\
\overleftarrow{\bm{h}} &=& \bm{\delta} - \bm{U} \otimes \bm{t}, \ \ \ 
\overrightarrow{\bm{h}} = \bm{\delta} - \bm{t} \otimes \bm{U},
\label{eq:Projection_U_G}
\end{eqnarray}
their siblings 
\begin{equation}
\overleftrightarrow{\bm{h}} = \overleftrightarrow{\bm{\gamma}}, \ \ \ \bm{h} = \bm{\gamma} - \bm{t} \otimes \underline{\bm{v}} - \underline{\bm{v}} \otimes \bm{t}
\label{eq:Projection_2_G}
\end{equation}
exist but cannot be expressed in terms of a spacetime metric or identity tensor.
(Note Equation~(\ref{eq:TimeForm}); and also, that while linear forms on $\mathbb{M}$ such as $\underline{\bm{U}}$ dual to vectors with time components do not exist on $\mathbb{G}$, the particular combination
\begin{equation}
-\frac{1}{c^2} \, \underline{\bm{U}} = \Lambda_{\bm{v}} \left( \bm{t} - \frac{1}{c^2}\, \underline{\bm{v}} \right) \rightarrow \bm{t}
\label{eq:TimeFormLimit}
\end{equation}
does limit sensibly as $c \rightarrow \infty$.
The index-lowered $\bm{v}$ is $\underline{\bm{v}} = \bm{\gamma} \cdot \bm{v}$.)
Due to the identity tensor in Equations~(\ref{eq:Projection_W_G}) and (\ref{eq:Projection_U_G}), information-preserving decompositions of $(1,1)$ tensors are possible on $\mathbb{G}$, while projections involving $(2,0)$ or $(0,2)$ tensors via Equation~(\ref{eq:Projection_2_G}) entail information loss.

\subsection{Mass Conservation on $\mathbb{G}$}

Decomposition of Equation~(\ref{eq:InertiaMomentumBalance}) on $\mathbb{G}$ for inertia-momentum balance provides an instructive example.
In the $c \rightarrow \infty$ limits of Equations~(\ref{eq:FourForce_U_M}) and (\ref{eq:FourStress_UU_U_M})--(\ref{eq:FourStress_UU_W_M}), the index-raised 4-force $\overrightarrow{\bm{\Upsilon}}$ and 4-stress $\overrightarrow{\bm{\Sigma}}$ lose their timelike components:
the index-raising $\overrightarrow{\bm{\Sigma}} = \bm{\Sigma} \cdot \overleftrightarrow{\bm{g}}$ on $\mathbb{M}$ that limits to $\overrightarrow{\bm{\Sigma}} = \bm{\Sigma} \cdot \overleftrightarrow{\bm{\gamma}}$ on $\mathbb{G}$ (and similarly for $\bm{\Upsilon}$) has a projective character that nullifies information on internal energy and heating.
Spacelike projections of {Equations}~(\ref{eq:InertiaMomentumBalance}) and (\ref{eq:InertiaMomentumTensor})   give the usual non-relativistic momentum balance.
However, the only timelike projection available on $\mathbb{G}$---contraction with $\bm{t}$---produces
\begin{equation}
\bm{\nabla} \cdot ( m \, \bm{N} ) = 0,
\end{equation}
which together with Equations~(\ref{eq:DerivativeProperTime}), (\ref{eq:BaryonConservation}), and (\ref{eq:ContinuumFourVelocity}) implies the conservation of mass $\mathrm{d}m/\mathrm{d}\tau = 0$ that held sway until Einstein.

Thus inertia of a continuum, represented in the $(2,0)$ tensor $\bm{T}$, has been decoupled from its energy in the passage from $\mathbb{M}$ to $\mathbb{G}$.
The apparent consequence, long assumed, has been that a complete picture of the energy of a continuum on $\mathbb{G}$ requires the first law of thermodynamics as an independent postulate (e.g., \cite{Ferrarese2008Introduction-to}).

%%%%%%%%%%%%%%%%%%%%%%%%%%%%%%%%%%%
\section{A More Unified View of Classical Mechanics on Minkowski and Galilei/Newton Spacetimes}
\label{sec:Unified}

While the inertia-momentum vector $\bm{I} = m \, \bm{U}$ of a particle exists on $\mathbb{G}$, the total energy-momentum form $\underline{\bm{I}} = m \, \underline{\bm{U}}$ does not because of the absence of a spacetime metric (note that the first and second equations of Equation~(\ref{eq:InertiaEnergyMomentumRepresentation}) do and do not make sense respectively as $c \rightarrow \infty$). 
Thus at first glance it looks as though the vector version of Newton's second law in Equation~(\ref{eq:ParticleNewtonSecondLaw_M}) can exist on $\mathbb{G}$, but the linear form version cannot.

However, information on internal energy and external heating need not be regarded as completely lost in the passage from $\mathbb{M}$ to $\mathbb{G}$. 
To motivate this I introduce the concepts of {``relative 4-velocity'' $\bm{V}$} and {``relative 4-momentum'' $\bm{P}$} of a particle as a vector and linear form respectively on both $\mathbb{M}$ and $\mathbb{G}$.
Give the relative 4-velocity the unified definition
\begin{equation}
\bm{V} = \bm{U} - \bm{w},
\label{eq:RelativeFourVelocity}
\end{equation}
which from Equations~(\ref{eq:FourVelocity_M}) and (\ref{eq:FourVelocity_G}) results in the more specific expressions
\begin{eqnarray}
\bm{V} &=& ( \Lambda_{\bm{v}} - 1 ) \, \bm{w} + \Lambda_{\bm{v}} \, \bm{v} \ \ \ (\mathrm{on \ } \mathbb{M}), 
\label{eq:RelativeFourVelocity_M} \\
\bm{V} &=& \bm{v} \ \ \ (\mathrm{on \ } \mathbb{G}).
\end{eqnarray}
Define the relative 4-momentum as
\begin{eqnarray}
\bm{P} &=& - m c^2 ( \Lambda_{\bm{v}} - 1 ) \, \bm{t} + m \, \Lambda_{\bm{v}} \, \underline{\bm{v}} \ \ \ (\mathrm{on \ } \mathbb{M}), 
\label{eq:RelativeFourMomentum_M} \\
\bm{P} &=& - \frac{1}{2} \, m \, \bm{\gamma}(\bm{v}, \bm{v}) \, \bm{t} + m \, \underline{\bm{v}} \ \ \ (\mathrm{on \ } \mathbb{G}),
\label{eq:RelativeFourMomentum_G}
\end{eqnarray}
where $\bm{\gamma} \cdot \bm{U} = \Lambda_{\bm{v}} \, \underline{\bm{v}}$ on $\mathbb{M}$ and $\underline{\bm{v}}$ on $\mathbb{G}$. 
While the relation $\bm{P} = m \, \underline{\bm{V}} = m \, \bm{g}~\cdot~\bm{V}$ on $\mathbb{M}$ (see {Equations}~(\ref{eq:TimeForm}) and (\ref{eq:RelativeFourVelocity_M})) does not exist on $\mathbb{G}$, Equation~(\ref{eq:RelativeFourMomentum_G}) is the perfectly sensible $c \rightarrow \infty$ limit of Equation~(\ref{eq:RelativeFourMomentum_M}). 
Thanks to the constancy of $\bm{t}$ on $\mathbb{M}$, the dynamical law 
\begin{equation}
\frac{\mathrm{d}\bm{P}}{\mathrm{d}\tau} = \bm{\Upsilon}
\label{eq:ParticleNewtonSecondLaw_M_G}
\end{equation}
for an ``elementary'' particle of constant mass $m$ is equivalent to the linear form version of Equation~(\ref{eq:ParticleNewtonSecondLaw_M}).
This equation also applies on $\mathbb{G}$, where thanks to {Equation}~(\ref{eq:TimeForm}) and (\ref{eq:TimeFormLimit})   the 4-force linear form limits to 
\begin{equation}
\bm{\Upsilon} = - \theta \, \bm{t} + \bm{f} = - \Theta \, \bm{t} + \bm{F}. 
\label{eq:FourForce_D_G}
\end{equation}
Thus on $\mathbb{G}$, contraction of Equation~(\ref{eq:ParticleNewtonSecondLaw_M_G}) with $\overleftarrow{\bm{\gamma}}$ gives Newton's second law
\begin{equation}
\frac{\mathrm{d}\bm{p}}{\mathrm{d}t} = \bm{F}
\label{eq:ParticleMomentum_G_31}
\end{equation}
in terms of the non-relativistic 3-momentum $\bm{p} = m \, \underline{\bm{v}}$.
Additionally, on $\mathbb{G}$, contraction of Equation~(\ref{eq:ParticleNewtonSecondLaw_M_G}) with $\bm{U}$ vanishes, and contraction with $\bm{V}$ or $\bm{w}$ gives the Work-Energy Theorem
\begin{equation}
\frac{\mathrm{d}e_{\bm{v}}}{\mathrm{d}t} = \bm{F} \cdot \bm{v}
\label{eq:ParticleEnergy_G_31}
\end{equation}
in terms of the particle kinetic energy $e_{\bm{v}} = m \, \bm{\gamma}( \bm{v}, \bm{v} ) / 2$.
(Beware that unlike their vector counterparts $\overrightarrow{\bm{F}} = \overrightarrow{\bm{f}}$ since $\overleftrightarrow{\bm{\gamma}} = \overleftrightarrow{\bm{h}}$ on $\mathbb{G}$, the linear forms $\bm{F} = \bm{\Upsilon} \cdot \overleftarrow{\bm{\gamma}}$ and $\bm{f} = \bm{\Upsilon} \cdot \overleftarrow{\bm{h}}$ are not equal!) 

The case of a material continuum is a straightforward generalization.
In its natural $(1,1)$ incarnation, the 4-stress of Equations~(\ref{eq:FourStress_UU_U_M}) and (\ref{eq:FourStress_UU_W_M}) limits {as $c \rightarrow \infty$} to the alternative decompositions
\begin{eqnarray}
\bm{\Sigma} &=& n \, \bm{U} \otimes \frac{\epsilon}{n} \, \bm{t}  + \bm{q} \otimes  \bm{t} + \bm{\varsigma}
\label{eq:FourStress_UD_U_G} \\
 &=& \bm{w} \otimes E \, \bm{t}  + \bm{Q} \otimes  \bm{t} + \bm{\sigma}
\label{eq:FourStress_UD_W_G}
\end{eqnarray}
on $\mathbb{G}$.
As a $(1,1)$ tensor, information on internal energy density and flux survive the $c \rightarrow \infty$ limit.
\mbox{A unified} 4-dimensional version of Newton's second law for an infinitesimal continuum element on $\mathbb{M}$ and $\mathbb{G}$, including both an external force and internal stresses, reads
\begin{equation}
n \, \frac{\mathrm{d}}{\mathrm{d}\tau} \left( \frac{\bm{\Pi}}{n} \right) = n \bm{\Upsilon} + \bm{\nabla} \cdot \bm{\Sigma}
\label{eq:ContinuumNewtonSecondLaw_M_G}
\end{equation}
with $\bm{\Pi} = n \, \bm{P}$. 
Thanks to {Equations}~(\ref{eq:DerivativeProperTime}), (\ref{eq:BaryonConservation}) and (\ref{eq:ContinuumFourVelocity})   this is equivalent to 
\begin{equation}
\bm{\nabla} \cdot \bm{S} = n \bm{\Upsilon},
\label{eq:RelativeFourMomentumBalance}
\end{equation}
where 
\begin{equation}
\bm{S} = n \, \bm{U} \otimes \bm{P} - \bm{\Sigma}
\label{eq:RelativeEnergyMomentumFluxTensor}
\end{equation}
is the $(1,1)$ {``relative energy-momentum flux tensor''}; compare Equation~(\ref{eq:InertiaMomentumTensor}).
On both $\mathbb{M}$ and $\mathbb{G}$, \mbox{contraction of} Equation~(\ref{eq:ContinuumNewtonSecondLaw_M_G}) or (\ref{eq:RelativeFourMomentumBalance}) with $\overleftarrow{\bm{\gamma}}$, $\bm{U}$, $\bm{V}$, and $\bm{w}$ respectively yield balance of 3-momentum, internal energy (first law of thermodynamics), kinetic energy (work-energy theorem), \mbox{and internal + kinetic energy} (in conservative form).
On $\mathbb{G}$, the first three respectively turn out to be the familiar non-relativistic relations
\begin{eqnarray}
n \, \frac{\mathrm{d}}{\mathrm{d}t} \left( \frac{\bm{\pi}}{n} \right) &=& n \, \bm{F} + \bm{D} \cdot \bm{\sigma}, 
\label{eq:MaterialMomentum_G_31} \\
n \, \frac{\mathrm{d}}{\mathrm{d}t} \left( \frac{\epsilon}{n} \right) &=& n \, \theta - \bm{D} \cdot \bm{q} + \bm{\sigma} : \bm{D} \bm{v},
\label{eq:MaterialInternalEnergy_G_31} \\
n \, \frac{\mathrm{d}}{\mathrm{d}t} \left( \frac{\epsilon_{\bm{v}}}{n} \right) &=& n \, \bm{F} \cdot \bm{v} + ( \bm{D} \cdot \bm{\sigma} ) \cdot \bm{v},
\label{eq:MaterialKineticEnergy_G_31}
\end{eqnarray}
where $\bm{\pi} = n \, \bm{p}$ is the 3-momentum density and $\epsilon_{\bm{v}} = n \, e_{\bm{v}}$ is the bulk kinetic energy density, while $\bm{\sigma}$ is the Cauchy 3-stress, defined here as a $(1,1)$ tensor field with components ${\sigma^i}_j$, so that $\bm{\sigma} : \bm{D} \bm{v} = {\sigma^a}_b \, D_a v^b$; and finally the contraction with $\bm{w}$ yields
\begin{equation}
\frac{\partial \epsilon_\mathrm{kin}}{\partial t} + \bm{D} \cdot ( \epsilon_\mathrm{kin}\, \bm{v} + \bm{q} - \bm{\sigma} \cdot \bm{v} )  = n \left(\theta + \bm{F} \cdot \bm{v} \right), 
\label{eq:FiducialEnergy_G_31}
\end{equation}
where $\epsilon_\mathrm{kin} = \epsilon + \epsilon_{\bm{v}}$.
This last equation, for bulk kinetic plus internal energy---that is, macroscopic and microscopic kinetic energy---also follows from the sum of Equations~(\ref{eq:MaterialInternalEnergy_G_31}) and (\ref{eq:MaterialKineticEnergy_G_31}), as obtained in the traditional approach when one regards Equations~(\ref{eq:MaterialMomentum_G_31}) and (\ref{eq:MaterialInternalEnergy_G_31}) as independent postulates on $\mathbb{E}^3$.

%%%%%%%%%
\section{Conclusions}

Greater conceptual unity of the relativistic and non-relativistic classical mechanics of material particles and continua is achieved by combining kinetic energy and 3-momentum in a linear form $\bm{P}$ (particles) or $(1,1)$ tensor $\bm{S}$ (continua) on Minkowski and Galilei--Newton spacetimes $\mathbb{M}$ and $\mathbb{G}$ (see Equations~(\ref{eq:RelativeFourMomentum_M}), (\ref{eq:RelativeFourMomentum_G}) for $\bm{P}$ and (\ref{eq:RelativeEnergyMomentumFluxTensor}) for $\bm{S}$).
Defining $\bm{P}$ as a linear form instead of as a vector geometrizes the deep principle that momentum is conjugate to displacement (a vector).
Additionally, as noted by Weyl \cite{Weyl1922Space---Time---}, it is natural that force be regarded as a linear form, so that direct contraction---without a scalar product---with displacement (or velocity) yields work (or power).

As on $\mathbb{M}$, this perspective allows the first law of thermodynamics to be regarded on $\mathbb{G}$ as a consequence of a unified dynamical law, Equation~(\ref{eq:ContinuumNewtonSecondLaw_M_G}) or (\ref{eq:RelativeFourMomentumBalance}), rather than an independent postulate.
Perhaps long familiarity with the luxury of a spacetime metric and insufficient attention to a thoroughly geometric perspective on the non-relativistic case have led to this possibility being long overlooked.

Nevertheless, this viewpoint retains some limitations inherent to the non-relativistic case.
\mbox{While the} total (internal + bulk) kinetic energy is governed by Equation~(\ref{eq:RelativeFourMomentumBalance}), inertia remains separate and is governed by the more familiar Equation~(\ref{eq:InertiaMomentumBalance}) with different implications on $\mathbb{M}$ and $\mathbb{G}$.
\mbox{Moreover the} definitions of the relative energy-momentum form and tensor $\bm{P}$ and $\bm{S}$, as indicated by the adjective ``relative,'' depend on the selection of a (family of) fiducial frames associated with $\bm{w}$ (cf. Equation~(\ref{eq:RelativeFourVelocity})).
Thus, while $\bm{P}$ and $\bm{S}$ are tensors on $\mathbb{M}$ and $\mathbb{G}$, as tensors defined in terms of a family of fiducial observers $\bm{w}$ their timelike components with respect to other frames do not manifest the non-relativistic transformation rule for kinetic energy.

In connection with this dependence of $\bm{P}$ (and therefore also $\bm{S}$) on the selection of $\bm{w}$, \mbox{it is} worth mentioning again two works cited in Section~\ref{sec:Introduction} regarding the non-relativistic case. 
\mbox{Exploring non-relativistic} covariance in four dimensions, Duval and K\"unzle found the internal 4-stress tensor $\bm{\Sigma}$, the part of $\bm{S}$ that does not depend on a choice of reference observer $\bm{w}$ (see Equations~(\ref{eq:FourStress_UD_U_G}) and (\ref{eq:RelativeEnergyMomentumFluxTensor})).
Exhibiting the transformation of the non-relativistic bulk kinetic energy requires an additional dimension, as emphasized by de~Saxc\'e and Vall\'ee \cite{de-Saxce2012Bargmann-group-,de-Saxce2016Galilean-Mechan,de-Saxce20175-Dimensional-T}; from the perspective of the present work, the extra dimension in effect allows for variation of the reference observer $\bm{w}$.
In fact, the $4 \times 4$ matrix gathering components of non-relativistic $\bm{S}$ appears in Chapter 12 of Ref.~\cite{de-Saxce2016Galilean-Mechan}, and as a submatrix of the $4 \times 5$ matrix of ``energy-momentum-mass-tensor'' components in Chapter 13 of that book. 
The present work exhibits $\bm{S}$ in geometric terms and motivates its existence on $\mathbb{G}$ as an instantly recognizable $c \rightarrow \infty$ limit of an easily understood tensor on $\mathbb{M}$ (see {Equations}~(\ref{eq:FourStress_UU_U_M}), (\ref{eq:RelativeFourMomentum_M}), (\ref{eq:RelativeFourMomentum_G}) and  (\ref{eq:FourStress_UD_U_G}) for expressions on $\mathbb{M}$ and $\mathbb{G}$, which enter a unified Equation~(\ref{eq:RelativeEnergyMomentumFluxTensor})).

Curved spacetime generalizations can be examined by allowing for non-constant fiducial fields associated with the $3+1$ foliation ($\bm{t}$ and $\bm{w}$ in the present work).

A final remark is that, as on $\mathbb{M}$ \cite{Ferrarese2008Introduction-to}, a free particle on $\mathbb{G}$ can be given a Hamiltonian but not Lagrangian formulation.
The free particle energy on $\mathbb{G}$ can be expressed $e_{\bm{v}} = \bm{P} \cdot \bm{U} = \bm{P} \cdot \bm{I} / m$, corresponding to the Hamiltonian
\begin{equation}
\mathcal{H} = \bm{P} \cdot \bm{w} + \frac{1}{m} \, \bm{P} \cdot \overleftrightarrow{\bm{\gamma}} \cdot \bm{P}
\end{equation}
yielding the expected canonical relation
\begin{equation}
\frac{\mathrm{d}\mathcal{X}}{\mathrm{d}\tau} = \frac{\partial\mathcal{H}}{\partial \bm{P}} = \bm{w} + \frac{1}{m} \, \bm{P} \cdot \overleftrightarrow{\bm{\gamma}} = \bm{U}.
\end{equation}
The absence of a corresponding Lagrangian formulation is signaled by $\mathrm{det} \left( \partial \mathcal{H} / \partial P_\mu \partial P_\nu \right) = \mathrm{det} \left( \gamma^{\mu\nu} \right) = 0$.
This is a reminder of the range of possibilities allowed by a symplectic view of physics on spacetime \cite{Souriau1970Structure-des-s,Souriau1997Structure-of-Dy}: there is more to life, and perhaps to nature, than Lagrangians on pseudo-Riemann manifolds.
\vspace{6pt}

\funding{This work was supported by the U.S. Department of Energy, Office of Science, Office of Nuclear Physics under contract number DE-AC05-00OR22725.}

%%%%%%%%%%%%%%%%%%%%%%%%%%%%%%%%%%%%%%%%%%
%\acknowledgments{In this section you can acknowledge any support given which is not covered by the author contribution or funding sections. This may include administrative and technical support, or donations in kind (e.g., materials used for experiments).}

%%%%%%%%%%%%%%%%%%%%%%%%%%%%%%%%%%%%%%%%%%
%\conflictsofinterest{Declare conflicts of interest or state ``The authors declare no conflict of interest.'' Authors must identify and declare any personal circumstances or interest that may be perceived as inappropriately influencing the representation or interpretation of reported research results. Any role of the funders in the design of the study; in the collection, analyses or interpretation of data; in the writing of the manuscript, or in the decision to publish the results must be declared in this section. If there is no role, please state ``The funders had no role in the design of the study; in the collection, analyses, or interpretation of data; in the writing of the manuscript, or in the decision to publish the results''.} 
\conflictsofinterest{The author declares no conflict of interest. The funders had no role in the design of the study; in the collection, analyses, or interpretation of data; in the writing of the manuscript, or in the decision to publish the results.}

%%%%%%%%%%%%%%%%%%%%%%%%%%%%%%%%%%%%%%%%%%%
%%% optional
%\abbreviations{The following abbreviations are used in this manuscript:\\
%
%\noindent 
%\begin{tabular}{@{}ll}
%MDPI & Multidisciplinary Digital Publishing Institute\\
%DOAJ & Directory of open access journals\\
%TLA & Three letter acronym\\
%LD & linear dichroism
%\end{tabular}}

%%%%%%%%%%%%%%%%%%%%%%%%%%%%%%%%%%%%%%%%%%%
%%% optional
%\appendixtitles{no} % Leave argument "no" if all appendix headings stay EMPTY (then no dot is printed after "Appendix A"). If the appendix sections contain a heading then change the argument to "yes".
%\appendix
%\section{}
%\unskip
%\subsection{}
%The appendix is an optional section that can contain details and data supplemental to the main text. For example, explanations of experimental details that would disrupt the flow of the main text, but nonetheless remain crucial to understanding and reproducing the research shown; figures of replicates for experiments of which representative data is shown in the main text can be added here if brief, or as Supplementary data. Mathematical proofs of results not central to the paper can be added as an appendix.

%\section{}
%All appendix sections must be cited in the main text. In the appendixes, Figures, Tables, etc. should be labeled starting with `A', e.g., Figure A1, Figure A2, etc. 

%%%%%%%%%%%%%%%%%%%%%%%%%%%%%%%%%%%%%%%%%%
\reftitle{References}

% Please provide either the correct journal abbreviation (e.g., according to the “List of Title Word Abbreviations” http://www.issn.org/services/online-services/access-to-the-ltwa/) or the full name of the journal.

\def\prd{Phys. Rev. D}


\begin{thebibliography}{999}
\providecommand{\natexlab}[1]{#1}

\bibitem[{Landau} and {Lifshitz}(1987)]{Landau1987Fluid-Mechanics}
{Landau}, L.D.; {Lifshitz}, E.M.
\newblock {\em Fluid Mechanics}, 2nd ed.;  Course of Theoretical
  Physics; Pergamon: Oxford, UK,  1987; Volume~6, 

\bibitem[{Ferrarese} and {Bini}(2008)]{Ferrarese2008Introduction-to}
{Ferrarese}, G.; {Bini}, D.
\newblock {\em {Introduction to Relativistic Continuum Mechanics}}; 
  {Lecture Notes in Physics}; Springer: Berlin/Heidelberg, Germany; New York, NY, USA, 2008; Volume 727.

\bibitem[{Gourgoulhon}(2013)]{Gourgoulhon2013Special-Relativ}
{Gourgoulhon}, E.
\newblock {\em {Special Relativity in General Frames}}; Graduate Texts in
  Physics, Springer-Verlag: Berlin/Heidelberg, Germany,  2013.

\bibitem[{Weyl}(1922)]{Weyl1922Space---Time---}
{Weyl}, H.
\newblock {\em Space---Time---Matter}, 4th ed.; Methuen \& Co.: London, UK, 1922.

\bibitem[Cartan(1923)]{Cartan1923Sur-les-variete}
Cartan, {\'E}.
\newblock {Sur les vari{\'e}t{\'e}s {\`a} connexion affine et la th{\'e}orie la
  relativit{\'e} g{\'e}n{\'e}ralis{\'e}e. (premi{\`e}re partie)}.
\newblock {\em \mbox{Ann. Sci.} Ecole Norm. Super.} {\bf 1923}, {\em 40},~325--412.

\bibitem[Cartan(1924)]{Cartan1924Sur-les-variete}
Cartan, {\'E}.
\newblock {Sur les vari{\'e}t{\'e}s {\`a} connexion affine et la th{\'e}orie de
  la relativit{\'e} g{\'e}n{\'e}ralis{\'e}e. (suite)}.
\newblock {\em Ann. Sci. Ecole Norm. Super.} {\bf 1924}, {\em 41},~1--25.

\bibitem[Cartan(1986)]{Cartan1986On-manifolds-wi}
Cartan, {\'E}.
\newblock {\em On Manifolds with an Affine Connection and the Theory of General
  Relativity}; Bibliopolis: Napoli, \mbox{Italy,  1986.}

\bibitem[{Toupin}(1957)]{Toupin1957World-invariant}
{Toupin}, R.A.
\newblock {World invariant kinematics}.
\newblock {\em Arch. Ration. Mech. Anal.} {\bf 1957}, {\em
  1},~181--211.

\bibitem[{Truesdell} and {Toupin}(1960)]{Truesdell1960The-Classical-F}
{Truesdell}, C.; {Toupin}, R.
\newblock {The Classical Field Theories}. In {\em {Principles of Classical
  Mechanics and Field Theory}}; {Encyclopedia of Physcs}; {Fl\"ugge}, S., Ed.; Springer-Verlag: Berlin/Heidelberg, Germany,  1960; Volume III/1,  pp.
  226--793.

\bibitem[{Trautman}(1965)]{Trautman1965Foundations-and}
{Trautman}, A.
\newblock Foundations and Current Problems of General Relativity. In {\em
  {Lectures on General Relativity}}; {Trautman}, A.; {Pirani}, F.A.E., {Bondi},
  H., Eds.; Prentice-Hall: Englewood Cliffs, NJ, USA,  1965; pp. 1--248.

\bibitem[{Trautman}(1966)]{Trautman1966Comparison-of-N}
{Trautman}, A.
\newblock Comparison of Newtonian and Relativistic Theories of Space-Time. In
  {\em Perspectives in Geometry and Relativity: Essays in Honor of V\'aclav
  Hlavat\'y}; Hoffmann, B., Ed.; Indiana University Press: Bloomington, IN, USA, 1966;
  Chapter~42, pp. 413--425.

\bibitem[{K{\"u}nzle}(1972)]{Kunzle1972Galilei-and-Lor}
{K{\"u}nzle}, H.P.
\newblock {Galilei and Lorentz structures on space-time: Comparison of the
  corresponding geometry and physics}.
\newblock In {\em Annales de l'IHP Physique Théorique, Section A}; {{1972};} %Please add the publisher and location.
 Volume {17}, pp.~337--362.

\bibitem[{Cardall}(2019)]{Cardall2019Minkowski-and-G}
{Cardall}, C.Y.
\newblock {Minkowski and Galilei/Newton Fluid Dynamics: A Geometric 3 + 1
  Spacetime Perspective}.
\newblock {\em Fluids} {\bf 2019}, {\em 4},~1.

\bibitem[{Duval} and {K\"unzle}(1978)]{Duval1978Dynamics-of-con}
{Duval}, C.; {K\"unzle}, H.P.
\newblock Dynamics of continua and particles from general covariance of
  Newtonian gravitation theory.
\newblock {\em {Rep. Math. Phys.}} {\bf 1978}, {\em
  13},~351--368.

\bibitem[{de Saxc\'e} and {Vall\'ee}(2012)]{de-Saxce2012Bargmann-group-}
{de Saxc\'e}, G.; {Vall\'ee}, C.
\newblock {Bargmann group, momentum tensor and Galilean invariance of
  Clausius-Duhem inequality}.
\newblock {\em Int. J. Eng. Sci.} {\bf 2012}, {\em
  50},~216--232.

\bibitem[{de Saxc\'e} and {Vall\'ee}(2016)]{de-Saxce2016Galilean-Mechan}
{de Saxc\'e}, G.; {Vall\'ee}, C.
\newblock {\em Galilean Mechanics and Thermodynamics of Continua}; John Wiley
  \& Sons, Inc.: Hoboken, NJ, USA,  2016.

\bibitem[{de Saxc\'e}(2017)]{de-Saxce20175-Dimensional-T}
{de Saxc\'e}, G.
\newblock 5-Dimensional Thermodynamics of Dissipative Continua. In {\em Models,
  Simulation, and Experimental Issues in Structural Mechanics}; Springer Series in Solid and Structural Mechanics; {Fr\'emond},
  M., {Maceri}, F., {Vairo}, G., Eds.; Springer: Cham, Switzerland,  2017; Volume~8,  pp. 1--40.

\bibitem[Cardall(2017)]{Cardall2017Relativistic-an}
Cardall, C.Y.
\newblock {Relativistic analogue of the Newtonian fluid energy equation with
  nucleosynthesis}.
\newblock \mbox{{\em \prd}} {\bf 2017}, {\em 96},~123014.

\bibitem[Souriau(1970)]{Souriau1970Structure-des-s}
Souriau, J.M.
\newblock {\em Structure des Syst{\`e}mes Dynamiques}; Dunod: Paris, France, 1970.

\bibitem[Souriau(1997)]{Souriau1997Structure-of-Dy}
Souriau, J.M.
\newblock {\em Structure of Dynamical Systems: A Symplectic View of Physics};
    Birkh{\"a}user: Boston, MA, USA,  1997; Volume 149.

\end{thebibliography}
\end{document}